\documentstyle[12pt]{article}
\begin{document}
\begin{titlepage}
\title{
{\bf Fast K system generator of pseudorandom numbers}
}
{\bf
\author{
   N.Z.Akopov,E.M.Madounts\\
   Yerevan Physics Institute,375036 Yerevan,Armenia\vspace{1cm}\\
A.B.Nersesian\\
Institute of Mathematics,375019 Yerevan,Armenia\vspace{1cm}\\
G.K.Savvidy\\
Physics Department,University of Crete,71409 Iraklion,Greece\\
and\\
Institut f\"ur Theoretische Physik,D-6000 Frankfurt am Main,
Germany\vspace{1cm}\\
W.Greiner\\
Institut f\"ur Theoretische Physik,
D-6000 Frankfurt am Main,Germany
}
}
\date{}
\maketitle
\begin{abstract}
\noindent

We suggest a new fast algorithm for the matrix generator of random
numbers which has been earliar proposed in \cite{savvidy1,akopov1}.
This algorithm reduces $N^{2}$ operation of the matrix generator
to $NlnN$ and essentially reduces the generation time. It also
clarifies the algebraic structure of this type of K system generators.

\end{abstract}
\thispagestyle{empty}
\end{titlepage}
\pagestyle{empty}

\vspace{.5cm}

In the articles \cite{savvidy1,akopov1} the authors suggested
the matrix generator of pseudorandom
numbers based on Kolmogorov-Anosov K systems
\cite{kolmogorov,anosov}. This systems are the most stochastic
systems, with nonzero entropy
\cite{kolmogorov,anosov,fomin,savvidy2,savvidy3}.
The properties of this new class of matrix
generators were investigated with the  criterion $\chi^{2}$ and
the  discrepancy $D_{N}$  in different dimensions. In all
cases it shows very good statistical properties \cite{akopov1}
(for another examples see also \cite{abramyan}).
Matrix generators based on
different ideas are proposed in \cite{niederreiter,grothe}.

In the recent article \cite {akopov2} this matrix generator
was  tested in the Monte- Carlo
simulations of the Ising model in one and two-dimensions
where we have defined control of the calculations.
The same simulation was carried  out  with  the RANECU
generator. As it was shown in this particular case the
quality of  those  generators
are nearly the same, but  the speed of  the  matrix  generator,
which we will call MIXMAX, is slower than that of
the RANECU generator.
It take place because the number of operations
of the MIXMAX generator is of order $N^{2}$,
where $N$ is the dimension of  the matrix.

The aim of this article is to suggest a
new fast algorithm, which
reduces $N^{2}$ operations of MIXMAX to
$N\ln N$ and allows to gain the generation time.
It also clarifies the algebraic structure of this type of
K system matrix generators.

The  essence of the algorithm is that the  almost $Toeplitz$
structure  of  the matrix generator
earliar proposed in \cite{savvidy1,akopov1}
rises up to $\it circulant$. The operations  with the circulant
matrix can be exchanged to the
calculations  with  the  diagonal  matrix. The  diagonalisation  is
performed  by the  direct  and  inverse
discrete $Fourier$ transformations and as it is well known, the fast
$Fourier$
transformation reduces $N^{2}$  operations to $N\ln N$.
At this point the new algorithm accelerates the  generation time.

Let us pass to the details of the algorithm. The  matrix  generator
is defined as \cite{savvidy1,akopov1},

$$X_{n+1} =A \cdot  X_{n}, (mod~1),\eqno(1)$$
where $A$ is $N \times N$ dimensional matrix \cite{akopov1}

$$ A = \left( \begin{array}{c}

        2,3,4,.......,N~~,1 \\
        1,2,3,.....,N-1,1 \\
        1,1,2,.....,N-2,1 \\
        ................. \\
        ................. \\
        1,1,1,...,2,3,4,1 \\
        1,1,1,...,1,2,2,1 \\
        1,1,1,...,1,1,2,1 \\
        1,1,1,...,1,1,1,1
\end{array} \right) \eqno(2)$$
and $X_{0}$  is an  initial vector, whose components must be irrational.
For example $X^{(i)}_{0} =1/ \sqrt {\pi+i}$ and $i=1,..,N$.
The $\it trajectory$ of the
$K$ system (1) $X_{0},X_{1},X_{2}....$ represents desired sequense of
the random numbers  \cite{savvidy1}. This approach allows a
large freedom in choosing of the matrices $A$ for the K system
generators and of the initial vectors \cite{savvidy1}.

Let's present the matrix (2) in the form $A=A_{1}-A_{2}$,~where

$$A_{1}= \left( \begin{array}{c}
   2,3,4,...,~ N~,N+1 \\
   1,2,3,...,N-1,~N~ \\
   1,1,2,..,N-2,N-1 \\
   ................. \\
   1,1,1,...,1,2,3,4 \\
   1,1,1,...,1,1,2,3 \\
   1,1,1,...,1,1,1,2
\end{array} \right) , A_{2}= \left( \begin{array}{c}
           0,0,0,..., 0~~,N  \\
           0,0,0,..., 0 ,N-1 \\
           0,0,0,..., 0 ,N-2  \\
            ................. \\
           0,0,0,...,0,0,0,4 \\
           0,0,0,...,0,0,1,3  \\
           0,0,0,...,0,0,0,2 \\
           0,0,0,...,0,0,0,1
\end{array} \right) . \eqno(3)$$

So $A_{1}$ is Toeplitz matrix and $A_{2}$ is almost zero matrix.
Let us  extend the matrix $A_{1}$  up to
$\it circulant$ $\hat{A_{1}}$ of the dimension $2N \times 2N$

$$\hat{A_{1}} = \left( \begin{array}{c}
         2,3,4,...,~N~,N+1,~1~,~1~,..,1 \\
         1,2,3,..,N-1,~N~,N+1,~1~,..,1   \\
         1,1,2,..,N-2,N-1,N,N+1,..,1 \\
         .................................   \\
         ................................. \\
         5,6,7,...,1~~,1~~,1~~,1~~,..,4 \\
         4,5,6,...,1~~,1~~,1~~,1~~,..,3 \\
         3,4,5,..,N+1,1~,1~,1~,..,2
\end{array} \right) . \eqno(4)$$
The following matrices

$$S_{1} = \left( E, 0 \right),~~~~~~
  S_{2}=  \left( \begin{array}{c}
          E \\
          0
  \end{array} \right) , \eqno(5)$$
where  $E$ is identity  matrix of the dimension $N \times N$,
$S_{1}$ is the matrix of the dimension $N \times 2N$ and $S_{2}$
has dimension $2N \times N$ allow to represent $A_{1}$ in the
form

$$A_{1} = S_{1} \hat{A_{1}} S_{2}.$$
Now the initial generator (1) can be written in the form,

$$X_{n+1} = (A_{1}~X_{n} - A_{2}~X_{n}), (mod~1), \eqno(1a)$$
where

$$A_{1}~X_{n}  = S_{1} \hat{A_{1}}  S_{2}~X_{n}. \eqno(6)$$
As it is known, the circulant matrix $\hat{A_{1}}$ can be
represented in the form:

$$\hat{A_{1}}= F^{-1} D F^{1}, \eqno(7)$$
where $D$ is a diagonal  matrix,  $F^{1}$ and $F^{-1}$ are
accordingly direct and inverse discrete Fourier
transformations

$$F_{n,m}= \left ( \frac{1}{\sqrt{2N}}exp~ \{~
\frac{2\pi i}{2N}(n-1)(m-1) ~\}
\right )^{2N}_{n,m=1}, \eqno(8a)$$
and

$$D = diag \left ( D_{1}.....D_{2N} \right ). \eqno(8b)$$
Diagonal elements $D_{1},..,D_{2N}$ of the matrix (4) are
counted by the formula:

$$D_{j}  = 2+3r_{j}+...+Nr^{N-2}_{j}+(N+1)r^{N-1}_{j} +
r^{N}_{j} +r^{N+1}_{j} +...+r^{2N- 1}_{j} ,\eqno(9)$$
where  $r_{j}  = exp(~\frac{2\pi i}{2N}(j-1)~)$, $j=1,..,2N$.
Taking into consideration  the  last  equation we  may  get  the
following  formulae:

$$D_{1}  = (N+5)N / 2  ,$$
$$D_{2}  = -D_{1}, $$
$$D_{j}  = \frac{(2+(-1)^{j-1})~(1-(-1)^{j-1})}{(1-r_{j} )} +
\frac{((-1)^{j-1}  ((N-1)r_{j} -N) + r_{j} )}{(r_{j} -1)^{2}},
\eqno(10)$$
where $j=3,..,2N$.
Now  the  algorithm  of  the  matrix  generator  is  realized
by the scheme:

$$X_{n+1}   = (S_{1}~ F^{-1}~ D~ F~ S_{2}~X_{n}  + A_{2}~X_{n} )~~
(mod~1) \eqno(1b)$$
The representation (1b) of the MIXMAX (1) completely solves
the problem.

This algebraic approach makes clear that at the "heart" of the
generator (1) \cite{savvidy1,akopov1} is the $Toeplitz$ matrix
$A_{1}$ in (3). Indeed one can check that the determinant of
$A_{1}$ is equal to one and the absolute values of all it`s
eigenvalues are different from one, therefore the matrix $A_{1}$
also represents equally good the K system generator

$$X_{n+1}   = A_{1}~X_{n} = S_{1}~ F^{-1}~ D~ F~ S_{2}~X_{n}$$
In fact the algebraic structure of this K system  generators become
more transparent in the sense that as we have seen one can embed the Toeplitz
matrix $A_{1}$ into  the
circulant of order $2N$ and decompose the last one into the sum
of the powers of the $\it basic$ permutation matrix,$\Omega$

$$\hat{A_{1}} = 2+3\Omega +4\Omega^{2}+...+(N+1)\Omega^{N-1} +
\Omega^{N}+\Omega^{N+1}+...+\Omega^{2N-1}, \eqno(9a)$$
where

$$\Omega =\left( \begin{array}{c}
         0,1,0,.....,0 \\
         0,0,1,.....,0   \\
         .............   \\
         ............. \\
         0,0,0,.....,1 \\
         1,0,0,.....,0
\end{array} \right) . \eqno(11)$$
This also clearly demonstrates the fact that the growth of the
matrix elements of the matrix $A$ \cite{savvidy1,akopov1}
in (2) and $A_{1}$ in (3) in the vertical direction produces reach
spectrum of eigenvalues (9),
the additional condition which is necessary
to produce $\it many-scale$
mixing of the system and to ensure a slower growth of the
discrepancy $D_{N}(A)$ \cite{savvidy1}.

It is also possible to modify the basic permutation matrix
$\Omega$ in (11) to get a class of K system generators with
very simple structure

$$A =\left( \begin{array}{c}
         0,~~1~,~~0~,.....,~~0 \\
         0,~~0~,~~1~,.....,~~0   \\
         .............   \\
         ............. \\
         0,~~0~~,~~0~~,.....,~~1 \\
         (-1)^{N+1},a_{1},a_{2},..,a_{N-1}
\end{array} \right) . \eqno(12)$$
In the last case it is easy to compute the characteristic
polynomial of $A$

$$\lambda^{N}-a_{N-1}~\lambda^{N-1}-...-a_{1}~\lambda + 1 =
0 \eqno(13)$$
and therefore it`s eigenvalues $\lambda_{1},...,\lambda_{N}$

$$\lambda_{1}\cdot \cdot \cdot \lambda_{N} =1$$
$$............$$
$$\lambda_{1} +...+ \lambda_{N}= a_{N-1} .\eqno(14)$$
This formulas allow to choose eigenvalues and then to
construct matrix $A$ for K system generators.

For example if N=4 and $a_{1}=0,~a_{2}=3,~and~a_{3}=0$,
then

$$\lambda_{1} =\sqrt {\frac{3+\sqrt{5}}{2}},~
\lambda_{2} = -\sqrt {\frac{3+\sqrt{5}}{2}},~ $$
$$\lambda_{3} =\sqrt {\frac{3-\sqrt{5}}{2}},~
\lambda_{4} =-\sqrt {\frac{3-\sqrt{5}}{2}}, \eqno(15)$$
with an additional simplectic structure of $A$.

To convince that this algorithm  produces the numbers
of the same "quality"
as before \cite{akopov1} we check this algorithm for the
matrix of the size $128 \times 128$. Bellow we present numerical
results  which were obtained for MIXMAX generator (1) realized though
the algorithm (1b) together with RANECU generator. We have
considered $\chi^{2}_{D}$ criterion when $D=1,..,5$

$$
\begin{array}{clcr}
                 MIXMAX & D &  \chi^{2} \\

               &        1    &  10.90 \\
                &       2   &   88.89 \\
                 &      3    &  992.896 \\
                 &      4    &  9941.92 \\
                 &      5    &  100012.0
\end{array}
\begin{array}{clcr}
               RANECU & D &  \chi^{2} \\

               &        1    &  7.625 \\
                &       2   &   96.7  \\
                 &      3    &  990.52  \\
                 &      4    &  9976.64  \\
                 &      5    &  100416.25
\end{array} $$
The programm which compares generation time of the MIXMAX and RANECU
shows that they work nearly with the same speed.

In conclusion the authors would like to thank F.James for interest
and H.Niederreiter for provision of
\cite{niederreiter,grothe}. This work was supported in part by the
Alexander von Humboldt Foundation.

\vfill
\end{document}